\documentclass{article}

\def\ben{\begin{equation}}
\def\een{\end{equation}}
\def\bena{\begin{eqnarray}} 
\def\eena{\end{eqnarray}}

\input amssym.def
\input amssym.tex
\begin{document}

\hfuzz=100pt
\title{The  metric and strong coupling limit
of the  M5-brane} 
\author{G W Gibbons
\\
D.A.M.T.P.,
\\Centre for Mathematical Sciences,
\\ Cambridge University, 
\\ Wilberforce Rd,
\\ Cambridge CB3 OWA,
\\ U.K.
\\and 
\\
P C West
\\Department of Mathematics,
\\King's College London,
\\The Strand.
\\London WC2R 2LS,
\\U.K.
}

\maketitle

\begin{abstract}
We find the analogue of the Boillat metric of Born-Infeld
theory for the M5-brane. We show that
it provides the  propagation cone
of {\sl all} 5-brane degrees. In  an arbitrary background field,
this cone never
lies outside the Einstein cone. An energy
momentum tensor for the three-form  is defined and shown to satisfy the
Dominant Energy Condition. 
The theory is shown to be well defined for
all values of the magnetic field but there is 
a limiting electric field strength. We
consider the strong coupling limit of the M5-brane
and show that the corresponding theory is conformally invariant
and admits infinitely many conservation laws. 
On reduction to the Born-Infeld case this agrees with the
work of Bia{\l}nicki-Birula.
 
\end{abstract}
\section{Introduction}

The behaviour of the
low energy states of the super-string theories in ten dimensions
are described by  supergravity theories
which are uniquely specified by their type of supersymmetry.
These theories satisfy a form of the
the  Equivalence Principle: the characteristics 
and hence the limiting propagation speeds
of all the fields, be they  the graviton, the gravitini,
p-form fields, scalars and spinors  are given 
universally by the light-cone of the
Einstein metric $g_{mn}$. The characteristics cones determine the
 paths of null geodesics
which are associated with massless particles..
Violations of the Weak   Equivalence Principal,
that is that all freely falling particles, massive or massless,
follow geodesics of the Einstein metric, 
are known to occur for massive particles in string theory,
because the former  can couple to a background dilaton field
$\Phi$.
This may be  seen by noting that the 
 unique effective actions, that is the maximal supergravity theories
in ten dimensions
\cite{IIA} \cite{IIB}, 
 contain different powers of $e^\Phi$
in front of the  field strengths associated with the 
Neveu-Schwarz$\otimes$Neveu-Schwarz and Ramond$\otimes$Ramond sectors.   
It is also known that string theory
induces higher derivative corrections which may 
affect the characteristics \cite{GibbonsRuback}
in a non-trivial background.

The behaviour of the  states of  open string theories
are described by Dirac-Born-Infeld type actions
provided one discards derivatives 
acting on field strengths.
In a 
recent paper \cite{GibbonsHerdeiro}
 it was argued that in a non-trivial
background 
these  open string states
obey  a modified
form of the Equivalence
Principle: the characteristics are universally given by
a metric first introduced in non-linear electrodynamics
 by Boillat
and which is conformal to what is usually referred to as the open
string metric \cite{Nappi}.

It was found that the Boillat cone never lies outside the Einstein cone
and in general touches it along the two principal null directions
of the two form ${\cal F}_{\mu \nu}=F_{\mu \nu}+ B_{\mu \nu}$,
where $F_{\mu \nu}$ is the Born-Infeld gauge field strength
 and $B_{\mu \nu}$ is the Kalb-Ramond 2-form gauge field.
In the limit of large field strength 
or equivalently large $\alpha ^\prime$,
propagation perpendicular to the principal null directions is suppressed. 

In this paper we shall  consider the analogue 
of the Boillat metric  for the M-5-brane equations of motion
\cite{Westcov}. We shall call this metric $C_{mn}$
Note that the characteristics
may be read of from the co-metric $C^{mn}$.
The speed of the rays are read off from its inverse $(C^{-1})_{mn}$
which is traditionally called a metric. If more than one metric
is involved one must  distinguish carefully between
metrics and their inverses, i.e. between metrics and  co-metrics.
In the context of lattices and elsewhere, the prefix ``co-''
is frequently replaced by the adjective ``reciprocal''
We prefer ``co-" because it is shorter. 
Note that there is no analogue of the
Boillat co-metric for the M-2-brane because the
world volume theory has no gauge fields.
We shall find that indeed a modified form of the Equivalence
Principal holds for the M-5-brane: the characteristics of 
fluctuations
are given universally by
an  analogue of the Boillat cone 
which never lies outside the Einstein cone.
We will show that if one 
dimensionally reduces the theory to Born-Infeld
theory then the fivebrane metric becomes  
the Boillat metric of the D4-brane. We   shall
define an energy momentum tensor for the fivebrane. It 
satisfies the  striking identity
\ben
T^{mn}=g^{mn}-C^{mn} \label{Hooke}
\een and we show that it obeys the dominant energy condition.

One may  show that the fivebrane has a limiting electric field strength 
beyond which the theory breaks down. By contrast, there is no 
limiting magnetic field strength. Finally we consider a 
high energy, or zero tension, limit of the fivebrane which is 
Weyl invariant and admits infinitely many conservation laws.

The plan of the paper is as follows. In section 2 we introduce the
5-brane metric and 5-brane Clifford algebra. We show that
the 5-brane metric  gives the characteristics of the
propagating fields whose speeds can never exceed that of light.
In section 3 we show that the two lightcones touch along a circle of
null directions.
In section 4 we introduce the covariantly conserved energy 
momentum tensor for the three-form
and show that it satisfies an identity analogous to
Hooke's law in classical elasticity theory. We show that in general
its trace is non-vanishing and hence the three-form  equations of
motion
are not invariant under a Weyl rescaling of the Einstein metric.
In section 5 we develop the idea that there is a 5-brane Equivalence
Principal.
In section 6 we show how our metric agrees with the Boillat metric
of Born-Infeld theory. 
In section 8  we show that plane wave
solutions of the linear theory are exact solutions of the full non-linear
theory.
In section 9 we discuss
the limiting electric field strength and the behaviour
 the theory near it. In section 10 we consider this strong coupling
limit in detail. We show that the theory is Weyl-invariant in this
limit
and that it admits infinitely many conserved quantities.

\section{ The 5-Brane Metric}

The  equations of motion of the 
scalars $X^N$, closed three-form, $H_{lmn}$ and spinors
$\Theta$  \cite{Westcov}
\ben
G^{mn}\nabla _m \nabla _n X^N=0,
\een
\ben
G^{mn}\nabla _m H_{nrs}=0
\een
\ben
\nabla _m\Theta (1-\Gamma) \Gamma^n m_{mn}=0 \label{Dirac},
\een
\noindent where $\nabla _m$ is the Levi-Civita covariant
derivative with respect to the Einstein metric $g_{mn}$. 
To begin with, we consider their
characteristics
which require only the leading derivative terms.
A short calculation shows that 
the characteristics of  the $X^N$ and
 $H_{lmn}$ fields    given  by the co-metric $G^{mn}$ 
which is defined  by
\ben
G^{mn}= m^{mr}g_{rs} m^{sn}= (1+{ 2\over 3} k^2 ) g^{mn} -4k^{mn},
\een
where $G_{mn}$ which will be defined to be
$ (G^{-1})_{mn}$, with $G^{mn}=(G)^{mn}$. 
All indices are raised and lowered using the Einstein
metric $g_{ab}$, which is taken to have signature $-1,+1,+1,+1,+1,+1$,
{\sl with the exception of } $G_{mn}$ which will be defined to be
$ (G^{-1})_{mn}$, with $G^{mn}=(G)^{mn}$.
In fact it seems to be possible to avoid explicit use
of the covariant form of the metric in all of the equations of motion.
Only the contra-variant form is required.   
\ben
m^{mn}=g^{mn}-2k^{mn},
\een
\ben
k^{mn}=h^n \\ _{ab} h^{mab},
\een
where $h_{abc}$ is a self-dual three-form 
\ben
h_{abc}= { 1\over 6} \epsilon_{abcdef}h^{def},
\een
and 
\ben
k^2= k_{ab} k^{ab}.
\een

In what follows  shall repeatedly use  the 
identities \cite{Westeom, Westcov}:
\ben
g_{ab}k^{ab}=0 \label{trace},
\een
and 
\ben
k^{ab}g_{bc}k^{cd}= { 1 \over 6} g^{ad}k^2={ 1\over 6} g^{ad} k_{ef}k^{ef}.
\label{square}\een
Thus, for example,
\ben
C^{mn}=m^m_am^{an}=2Q ^{-1}m^{mn}-g^{mn} \label{metric}.
\een

The self dual field $h_{abc}$ obeys the relations  \cite{Westeom}
\ben
h_{abe} h^{cde}= \delta_{[a}^{[c} k_{b]}^{d]}, \label{kronecker}
\een
and  
\ben k_a{}^c k_c{}^b= {1\over 6}\delta_a^b k^2.
\een
For later use we define
\ben
Q=1-{ 2\over 3} k^2.
\een

 The closed three form $H_{abc}$ is related to the self-dual
field 
$h_{abc}$ by $h_{abc}=m_a{}^e H_{ebc}$, or equivalently by 
$H_{abc}=(m^{-1})_a{}^e h_{ebc}$ where $m^{-1}=Q^{-1}(1+2k)$. It
will be prove useful  in what follows to  translate the
self-duality condition on $h_{abc}$ to one expressed entirely in terms of
$H_{abc}$. This was carried out in \cite{Westeom} and refined in
 \cite{SezginSundell}. Since 
$k_a{}^e h_{ebc}$ is anti-self-dual we can express 
\ben 
H^+_{abc}\equiv {1\over 2}(H_{abc}+{1\over 3!}\epsilon_{ abcdef}
H^{def})=Q^{-1}h_{abc}\een

 and 
\ben
H^-_{abc}\equiv {1\over 2}(H_{abc}-{1\over 3!}\epsilon_{ abcdef}
H^{def})=2Q^{-1}k_a{}^e h_{ebc}\een

Taking the sum and difference we find that 
\ben
H_{abc}= Q^{-1}(1+2k)_a{}^e h_{ebc},\ 
* H_{abc}= Q^{-1}(1-2k)_a{}^e h_{ebc}
\een
Multipying the second equation by the matrix  $(1+2k)^2$  and using the first
equation we conclude that 
\ben
*H^a\thinspace _{bc}=Q^{-1}G^{ae} H_{ebc} \label{A}
\een

Substituting the relation $h_{abc}=m_a{}^e H_{ebc}$ into equation (\ref{kronecker}) 
we find that 
\ben
H_{abe}H^{cde}={1\over 2}\delta_{[a}^c\delta _{b]}^d
Q^{-1}(Q^{-1}-1)+2Q^{-2}k_{[a}^c k_{b]}^d + 
Q^{-2}(2-Q) \delta_{[a}^{[c} k _{b]}^{d]}
\label{B}\een
Taking the trace of this equation we find that 
\ben
k_a{}^c= {Q^2\over (2-Q)}((H^2)_a{}^c -{1\over 6} \delta _a{}^c H^2)
\een
 
and tracing again \ben
H^2=6 Q^{-1}(Q^{-1}-1)
\een
where $(H^2)_a{}^c= H_{aef}H^{cef}$.

We may express the last equation as 
\ben
Q= -{3\over H^2}(1-\sqrt{1+{2\over 3}H^2})
\een

It is now straightforward to express the matrix $m^2$ as 
\ben
(m^2)^{ac}= G^{ac}= {Q^2 \over (2-Q)}(\eta^{ac}(1+
{4\over 3}H^2 )-4(H^2)^{ac}) \label{easymetric1}
\een
 
and 
\ben
(m^{-2})_{ac}= G_{ac}= {1\over (2-Q)}(\eta_{ac}+
4(H^2)_{ac}) \label{easymetric}
\een

Finally, we may express the self-duality  
of $h_{abc}$ in terms of $H_{mnp}$ by using equations 
(\ref{A}) and (\ref{B})  we find that 
\ben
*H_{abc}={1 \over \sqrt {1+{2\over 3}H^2}}
((1+{4\over 3}H^2 )\delta _a{}^e-4(H^2)_a{}^e)H_{ebc} \label{easydual}
\een

We now turn briefly to fermion sector.
 The projector in the spinor equation of motion is defined by
\ben
\Gamma= -{ 1\over 6} \eta^g  _{lmnpqr} \Gamma ^{lmnpqr}+ { 1\over 3}
h_{lmn} \Gamma^{lmn}.
\een
where $\eta^g _{lmnpqr}$ is the alternating tensor (not density)
constructed from the Einstein metric.

In the case that the scalar and spinors vanish, the 
covariant equations
of motion, when expressed in terms of 5-dimensional language,
agree with a particular case of the equations of \cite{Perry}.
The particular case is the one that upon reduction gives 
Born-Infeld theory. One advantage of the covariant formulation 
used here is that, not only does it cover the 
more general case 
of non-vanishing scalars and spinors, but also
the derivation of the characteristics is especially
transparent.

The characteristics determine a metric only up to 
a conformal factor. It turns out that
it is more convenient to Weyl rescale the co-metric $G^{mn}$ and 
we therefore adopt as our definition of the M5-brane metric 
$C^{mn}$
\ben
C^{mn} = Q^{-1} G^{mn}= Q^{-1} m^{mp} g_{pq} m^{qn}.
\een

We recall that in the Born Infeld theory the characteristics 
are given by the Boillat metric\cite{GibbonsHerdeiro} 
up to a conformal factor. The Boillat metric which is 
proportional to the open string metric has the advantage that 
it is invariant under electric magnetic duality rotations 
rather than merely being invariant up to a conformal factor 
as is the open string metric. 
In a two  recent papers \cite{dutch} an analogue for the fivebrane 
of the open string metric was proposed. 
Specifically, it was suggested that  the analogous metric 
should be given by 
\ben 
\phi(H_{lpq}H^{lpq}) (g_{mn}+4H_{mpq}{H_n}^{pq}),   
\een
where the function $\phi$ should behave like 
$(H_{lpq}H^{lpq})^{-{2\over3}}$
for 
large $H_{lpq} H^{lpq}$ 
in order that, upon reduction,  it agree with the open 
string metric of string theory in the relevant limit. 

The two proposals are both 
conformally related to the metric in the fivebrane equations. 
We will see that with our choice of conformal factor we 
obtain the Boillat metric of the D4-brane upon reduction. 
Our choice has the additional advantage that in terms of it 
the equations of motion can be rewritten in  a natural 
way.

We now turn to the characteristics
 associated with the Dirac equation (\ref{Dirac}).  
While the  gamma
 matrices $\Gamma^m$ give a Dirac square root of the
restriction  of the bulk  Einstein co-metric to the brane,
\ben
\Gamma ^m \Gamma ^n + \Gamma ^n \Gamma ^m= 2 g^{mn},
\een
the spinor  equation of motion contains the 5-brane  Gamma
matrices 
${\tilde \Gamma} ^m= n^m\thinspace _n \Gamma^ n$, where $n^m\thinspace _n= 
Q^{- {1\over 2}} m^m\thinspace _n$  
which give a Dirac square root of the 5-brane co-metric $C^{mn}$
\ben
{\tilde \Gamma}^m{\tilde \Gamma}^n + {\tilde \Gamma}^n{\tilde \Gamma}^m=
2 C^{mn}.
\een

One may view the $n^m\thinspace _n$ as a sort of sechbein for the 
5-brane metric $C_{mn}$ since
\ben
C^{mn}= n^m \thinspace _p n^n \thinspace _q g^{pq}.
\een
Note that $n_{nm}= g_{ml}n^l \thinspace _m$ is symmetric.

Writing out the Dirac equation in terms of the gamma matrices ${\tilde
  \Gamma}^m
$ reveals that the spinor characteristics are also given by $C^{mn}$.

In the absence of a background 
$H_{lmn}$ field, 
 $C^{mn}$ and $g^{mn}$
coincide.  Note that there is a {\sl single} $G^{mn}$
and thus a single characteristic cone. 
That is just as in Born-Infeld theory,
there is no bi-refringence:
all  polarisation states travel with the same speed.
Since any non-linear electrodynamic theory, including
ones exhibiting bi-refringence can be made $N=1$
supersymmetric, its absence cannot be attributed
to just one supersymmetry. However one might imagine
that this property is a consequence of maximal superymmetry.
In the case of Born-Infeld theory, the absence of bi-refringence
, and the exceptional
property that the system exhibits no shocks,
characterises the theory uniquely (see \cite{GibbonsHerdeiro}
for references). It is an attractive conjecture that the same
uniqueness property holds for the M-5-brane equations of motion.

We now  establish that the 5-brane co-cone
$C^\star_G \in T^\star M$ lies outside or on the Einstein
co-cone $T^\star M \supset C_G ^\star  \supseteq C_g^\star$.
The notation here is as follows.
$C^\star _g$ consists of timelike co-vectors $p_m \in T^\star M$
 such that $g^{mn} p_m  p_n \le 0$
and its boundary  consists of the lightlike co-vectors $l_m$
for
which $g^{mn}l_m l_n=0$.
Since
passing to the dual space $TM$  reverses inclusions we have that the
5-brane  cone lies
inside or on the Einstein cone, $TM \supset C_g \supseteq C_G$.
In plain language this means that 5-brane excitations travel
with speeds no greater than that of light.
Note that the cone $C^\star _G$  depends only on the conformal
equivalence class of co-metrics of which $G^{mn}$
is  one representative and $C^{mn}$ another.

To establish our  basic causality  result we consider a co-vector $l_a$ lying
on the boundary of $C^\star_g$, i.e. for which
\ben
g^{mn}l_ml_n=0.
\een
    
Using (\ref{metric}) we find that 
\ben
C^{mn}l_ml_n=-4Q s_{mn}s^{mn},
\een
where $s_{mn}=-s_{nmb}=h_{mnl}l^l$. Thus \ben s_{mn}l^m=0.\een
By choice of frame $l^m=(1,0,0,0,0,1)$ and this $s_{05}=0$ and 
$s_{0i}+s_{5i}=0$, where $i=1,2,3,4$ are spatial indices. One thus has
\ben
s_{ab}s^{ab}=s_{ij}s^{ij}\ge 0.
\een

Thus
\ben
C^{mn}l_ml_n \le 0.
\een
It follows that the boundary of $C^\star_g$ lies inside or on $C^\star_G$
and we are done.

\section{ Principal Null Directions}

In the case of Born-Infeld theory in four spacetime dimensions, 
generically the Boillat
and Einstein cones touch along two principal null directions
of the electromagnetic two form $F_{\mu \nu}$ \cite{GibbonsHerdeiro}.
In fact this result holds in all dimensions.
The common null direction $l^m=(1, n_i)$ with
$n_in_i=1$ must satisfy
\ben
l^nF_{na} F_{mb}g^{ab} l^m=0.
\een

For a generic two-form one may find a frame in which
the only non-vanishing components are
$F_{01}=-F_{10}, F_{23}=-F_{32}, F_{45}=-F_{54} \dots$.
The touching condition becomes
\ben
F_{01}^2(1-n_1^2) + F^2_{23} (n_2^2 + n_3^2) + F^2_{45} (n_4^2 + n^2_5) \dots =0.
\een
There are  only two solutions ;  $l^m=(1,\pm 1,0, \dots ,0)$.

One may ask what is the analogue 
of this result
for the fivebrane with its self-dual three-form $h_{lmn}$
in six spacetime dimensions?
In order to answer this question we need to put
$k_{ab}$ rather than $F_{ab}$ in standard form by 
diagonalizing with respect to to $g_{ab}$.
Using (\ref{trace}) and (\ref{square})
one easily sees that generically
$k_{ab}$ takes the form, up to permutations
of the spatial axes,
\ben
k_{ab}= \sqrt {k^2 \over 6} {\rm diag}( 1,1,1,1,-1,-1).
\een

The common null directions must be common solutions of

\ben
-1+
n_1^2 +n_2^2 +n_3^2 +n_4^2 + n_5^2=0, 
\een
and 
\ben
1+n_1^2 +n_2^2 +n_3^2 -n_4^2 - n_5^2=0. 
\een
There is in general a circle  of such directions
\ben
l^m=(1,0,0,0,\cos \alpha , \sin \alpha),
\een
along which the Einstein and Boillat cones coincide.
In all other directions the Boillat cone lies inside
the Einstein cone.

\section{The Energy Momentum Tensor and Hooke's Law}

Following the discussion of the energy momentum tensor in 
\cite{Westemt},
we shall in this paper define the energy momentum tensor
as 
\ben
T^{mn}= g^{mn} - { G^{mn} \over Q}\label{tensor}. 
\een
The main difference from the 
energy momentum tensor defined in 
reference \cite{Westemt},also used in \cite{SezginSundell},
is that we have added the metric tensor $g_{mn}$ 
 to the tensor defined there 
so as to make the energy momentum tensor 
vanish for zero three-form field  $h_{abc}=0$.
In fact it is (\ref{tensor}) which enters directly into the 
superysymmetry algebra
and hence the Bogomol'nyi bound of the theory\cite{Westcal}.
The energy momentum tensor so defined  has some 
important positivity
properties which we will explore shortly.

In terms of $C^{mn}$  we have 
the strikingly simple formula
\ben
T^{mn}=g^{mn}-C^{mn} \label{hooke}
\een
This formula has an interesting  interpretation
which appears to be closely related to
Hooke's law in the classical theory of non-linear elasticity theory.
This is formulated in terms of diffeomorphisms $\phi$ 
from the  manifold $\Sigma$
of the elastic body into  flat
three-dimensional Euclidean space ${\Bbb E}^3$ with metric $\delta$. 
The relaxed or unstretched  state of least energy
corresponds to a diffeomorphism  $\phi_0$.
One defines the strain tensor
tensor   $\sigma_{ij}$ of a general 
stretched  state corresponding to
a diffeomorphism $\phi$
by
\ben
\sigma_{ij}= {\delta_\star}_{ij}-{\delta _0}_{ij}
\een
where 
$\delta_\star$ is the pull-back
of the flat Euclidean metric $\delta$ under the diffeomorphism $\phi$
and
$ {\delta_0}_{ij}$ is the pullback of the flat metric under $\phi_0$.
For an isotropic medium Hooke's law states  that the strain
$\sigma_{ij}$  is proportional to the applied stress
$T_{ij}$. Our formula (\ref{Hooke}) is similar
but not identical because  the tensors are contra-variant not co-variant. 
Thus  our case the analogue of  ${\delta_\star}_{ij}$ 
is the pull-back of the bulk
closed  string co-metric to the M-5-brane world volume.
The analogue of the unstretched metric ${\delta_0}_{ij}$
is the  5-brane co-metric $C^{mn}$.
We believe that it would
be worthwhile exploring
this analogy further. 

Another formula which is similar to one occurring
 Born-Infeld case \cite{GibbonsHerdeiro} is the remarkable identity 
\ben
\det C^{mn}=\det g^{mn} \label{det}.
\een

To check conformal invariance  we compute  the
trace of the energy momentum tensor. It is  given by

\ben
T_m^m= -{8k^2   \over ( 1-{2 \over 3}k^2) }
\een

Thus the theory is
Weyl-invariant in the weak field limit
in which the equations of motion become linear.
However it is not Weyl-invariant for finite values
of the fields.
We see shall see later that Weyl invariance is restored
in the strong coupling limit.

Later we shall  make use of some other identities 
involving the energy momentum tensor which we derive here.
Because the energy momentum tensor is conserved
with respect to the Levi-Civita connection
\ben
\nabla _m T^{mn}=0,
\een
we have, from Hooke's Law, the following identities
\ben
\nabla_n C^{mn}=0 \label{divergence}.
\een
  
These will be used in the next section to establish 
the M-5-brane Equivalence Principal.
Some additional useful identities 
may be obtained as follows.
One defines
\ben
n^{mn}= Q^{-{ 1\over 2}} m^{mn}=
 Q^{-{1 \over 2}}( g^{mn}-2k^{mn})
\een
Thus
\ben
n^{-1}_{mn}= Q^{-{ 1\over 2}} (g_{mn} +2k_{mn}).
\een
We thus have
\ben
C^{mn}= n^{ma} g_{ab} n^{bn},
\een
and of course
\ben
{\tilde \Gamma} ^n= n^{na}g_{ab} \Gamma^b.
\een
Now one easily finds
\ben
g_{mn}C^{mn}= 
{ 1+ { 2\over 3} k^2 \over 1- { 2\over 3} k^2 }=g^{mn}C_{mn} 
\een

In reference \cite{Westemt} an energy momentum 
tensor was introduced 
that treated the scalars and three form in a more symmetric fashion. 
This tensor is  given by 
\ben
 S^{mn}= T^{mn}-g^{mn} .
\een
and is covariantly conserved. 
Reference \cite{Westemt} also found a tensor density 
\ben
\frak{S} ^{mn}= \sqrt{-g}( T^{mn}-g^{mn} ).
\een
that was conserved in the sense 
\ben
\partial _m \frak{S} ^{mn}=0. 
\een

In the case that the three-form vanishes,
we have $T^{mn}=0$, but $S  ^{mn} \ne 0$,
and so the quantity $\sqrt{-g} g^{mn}$
is a measure of the energy-momentum of  the scalars.
In fact it is the canonical stress tensor 
with respect to the flat metric $\delta_{mn}$.
It may be obtained from the Lagrangian for the scalars
 written down in "static ",
or more accurately "Monge", gauge. 

\section{ The M5-Brane Equivalence Principle }

It is striking fact that  the 
co-metric $C^{mn}$ rather than the Einstein co-metric
$g^{mn}$ enters in  the equations of motion
of all the 5-brane fields in such a way that
it is impossible to determine the Einstein metric
 by means of observations
using 5-brane fields alone.
In this respect the situation resembles attempts
to reinterpret General Relativity as a flat space theory
 by introducing
a flat metric $\eta_{\mu \nu}$ 
into spacetime. The problem is that by the Equivalence Principal
no  physical measurement can detect the flat metric.

Our aim in this section is to explore further this  enhanced 
version of the 
Equivalence Principle for the fivebrane. 
Our claim is that  there exists a
 preferred set of variables to describe the theory 
that are related in a direct way to the physical observables. 
In particular, in section two we showed that the characteristics of the 
fivebrane were given by the metric $C_{mn}$. 
This means  that the metric can be determined up to a conformal factor 
by observing the motion of small fluctuations. Another physically relevant 
variable is the gauge invariant field strength $H_{lmn}$. 
This satisfies the Bianchi identity and hence can be written in terms of 
a two form potential that couples directly to a two brane probe. 
Hence using   a two brane probes allows one to measure the  $H_{lmn}$ field. 

To illustrate this point we now show how to write  the fivebrane  equations of 
motion entirely in terms of the variables $C_{mn}$ 
and $H_{lmn}$ and $X^N$.. In particular,  we will find that the 
equations of  motion for the scalar and the three-form can be written 
using the  Levi-Civita covariant derivative with respect to the 
metric $C_{mn}$. The situation  is  analogous to how the usual
Equivalence Principal works in General Relativity.
Because all  physical equations are writen in terms of the metric
$C_{mn}$ the metric $g_{mn}$ is not directly observable.

We begin with the scalar equation of motion.
Because  $G^{mn}$ is proportional  to $C^{mn}$,
this equation can be written as 
\ben
C^{mn} \nabla _m \nabla _n X^N=0.
\een
Following \cite{Westemt}  and using 
(\ref{divergence}) we may re-write this as
\ben
\nabla _m (C^{mn} \partial_n X^N)= { 1\over \sqrt{-g}}
\partial _m ( \sqrt{-g} C^{mn} \partial _n X^N )=0.
\een
Now using (\ref{det}) we have
\ben
{ 1\over \sqrt{-C} } 
\partial _m (\sqrt{-C} C^{mn} \partial _n X^N )=
\Delta _m ( C^{mn} \Delta _n X^N )=0 \label{Dalembert},
\een
where $\Delta _m$ is the Levi-Civita
covariant derivative with respect to the 5-brane metric $C_{mn}$
which is defined  by
\ben
\Delta_m C^{ab}=0.
\een
 We see that (\ref{Dalembert})
is just the covariant wave equation with respect
to the 5-brane metric $C_{mn}$.

The closure condition for the
three-form $H_{lmn}$  
\ben
\partial _{[p}H_{qrst]}=0,
\een
clearly requires no metric.
The equation of motion of the three-form
may  be written
as
\ben
C^{mn} \nabla _m (H_{nab})=0. \label{field}
\een
Using (\ref{divergence}) and the fact that $\nabla _mg^{ab}=0$,
we re-write (\ref{field}) as
\ben
\nabla _m ( P^{mab})= 0,
\een
where
\ben
P^{mab}= C^{mn}g^{ac} g^{bd} H_{ncd}.
\een
Now the contravariant tensor $P^{mab}$ is
totally antisymmetric and satisfies (\cite{Westemt} equation(17))

\ben
P^{lmn}=\star_g H^{lmn}
\een
and  therefore 
we may rewrite (\ref{field}) 
as
\ben
{ 1\over \sqrt{-g}} \partial _m ( \sqrt {-g} P^{mab} )=0.
\een
Now using (\ref{det}) we get
\ben
{ 1\over \sqrt{-C}} \partial _ m ( \sqrt{-C} P^{mab} )=0. 
\een
This may now be put in the fivebrane metric  covariant form
\ben
\Delta _m(\star_C H^{mnp})=0.
\een
  
The Hodge operations $\star_g$ and $\star_C$ are taken with respect
to the Einstein and 5-brane metric respectively.
However they are related because if  $\eta_C^{mabncd}$ is the contravariant
alternating tensor which is covariantly constant
with respect to the Levi-Civita connection of the 5-brane
metric $C_{mn}$, we have
\ben
\sqrt {-C} \eta_C^{mabncd}= \sqrt{-g} \eta_g^{mabncd},
\een
 where $\eta_g^{mabncd}$ is the contravariant
alternating tensor which is covariantly constant
with respect  to the Levi-Civita connection of the 
Einstein  metric $g_{mn}$.

Finally we consider the Dirac equation. Following \cite{Westemt} 
 equation (13), this may be written as  
\ben
\nabla _m( \Psi ^\prime (1- \Gamma) {\tilde \Gamma} ^m )=0,
\label{bar}
\een
where
\ben
\Psi ^\prime = Q^{-{ 1\over 2}} \Theta. \label{P}
\een

Now we must rewrite the projector $\Gamma$
in terms of the 5-brane gamma-matrices ${\tilde \Gamma} ^m$.
We  have the formula
\ben
\Gamma = -{ 1\over 6} {\eta_ C}_ {lm
nrst} {\tilde \Gamma }^l {\tilde \Gamma }^m {\tilde \Gamma}^n
{\tilde \Gamma }^r {\tilde \Gamma }^s {\tilde \Gamma}^t + { 1\over 2}
H_{lmn}{\tilde \Gamma }^l {\tilde \Gamma }^m {\tilde \Gamma }^n.
\een
The covariant derivative $\nabla _m$ is the spinor 
derivative with respect to the Einstein metric. We expect that 
one may be able to rewrite this 
in terms of the spinor covariant derivative $\Delta _m$. This is likely to 
require  a more elaborate redefinition of the spinor 
than in   (\ref{P}), 
 along the lines of that discussed in reference 
\cite{Westcal}.

\section{Dimensional Reduction and the relation to the 
Open String metric}

It is known \cite{Westeom, Westcov} that under double dimensional reduction
to five-dimensions the equations of motion of the
five brane reduce to those of Dirac-Born-Infeld
theory. We shall
now investigate the relation between
the fivebrane metric and the Boillat metric
of the D4-brane  in this case.
We assume that  the fields $h_{lmn} $
are independent of the fifth spatial dimension and that
\ben
H_{mn5}=F_{mn},
\een

Using the results of \cite{Westcov}
eqn(136) \cite{Westcov} we have (after rescaling $F$ to be consistent with
the standard normalization)
we find the reduction of five-brane metric to be given by

\ben
C^{mn}= Z (1-F^2)^{-1}
\een
where
\ben
Z=\sqrt{1+2x-y^2}= \sqrt{-\det(g_{mn}-F_{mn})}.
\een

The 
Boillat co-metric is given by
\ben
C_{\rm Boillat}^{mn}= Z \Bigl ((1-F^2)^{-1}\Bigr )^{mn}
\een

In other words the two metric coincide.

We now extend the work of \cite{GibbonsHerdeiro}
to give 
an enhanced Equivalence Principle similar to that discussed
for the fivebrane earlier.
The Born-Infeld equations may be written as
\ben
\partial _{[l} F_{mn ]}=0, \label{Faraday}
\een
and
\ben
\nabla _m P^{mn}= { 1\over \sqrt{-g} }\partial _m (\sqrt {-g} P^{mn})=0.
\label{Ampere1}
\een
Clearly (\ref{Faraday}) requires no metric and (\ref{Ampere1})
may be re-written (using (\ref{det}))
in terms of the Boillat metric  as. 
\ben
\Delta  _m P^{mn}= { 1\over \sqrt{-C} }\partial _m (\sqrt {-C} P^{mn})=0.
\label{Ampere2}.
\een

One may check, just as for the 5-brane,
that the scalar equations of motion
may be written as 

\ben
\Delta_m \Delta ^m X^N =0.
\een

\section{ Dominant Energy Condition}

Almost all physically well behaved classical energy momentum tensors
satisfy the Dominant Energy Condition. This states that for
every pair of future directed casual vectors $p^m,q^m \in C^+_g$
one has
\ben
T_{mn}p^m q^n \ge 0 \label{dominant}.
\een

Note that, if the Dominant Energy Condition
holds with respect to the Einstein metric,
it necessarily holds with respect to the
fivebrane metric, since the lightcone of the former
includes that of the letter.

The importance of the Dominant  Energy Condition
is that it guarantees  the classically
fields whose energy momentum tensors satisfy
the condition  propagate  causally.
 It is also  an essential ingredient in
the Positive Energy Theorems
of General Relativity. A theorem of Hawking
implies that if the Dominant Energy Condition holds
then  matter cannot escape from,
or enter,  a bounded spatial region  
at a speed faster than light \cite{Hawking}. In particular it
guarantees some sort of stability
since  matter obeying the  condition cannot just 
simply disappear
\cite{Hawking}.  

Let us evaluate the left hand side of (\ref{dominant}) for the fivebrane.
Using equations (\ref{metric},  \ref{Hooke}) we  find that 
\ben
Q^{-1}( 2k_{mn}p^m q^n+ { 2 \over 3} k^2   p \cdot q), 
\een
where 
\ben
p\cdot q=-g_{mn}p^mq^n \ge0.
\een

We will now show that the left-hand-side of the above equation 
is indeed positive since the quantities $k^2$ and 
$k_{mn}p^mq^n$and are non-negative.

To see that $k^2$ is positive we introduce an electric 
two-form in five dimensions by
\ben
e_{ij}=h_{oij}.
\een
Self-duality   of $h_{lmn}$ implies that
\ben
e_{ij}=b_{ij}=- { 1\over 6}\epsilon_{ijkrs} h^{krs}
\een

Calculation 
reveals that

\ben
k_{ij}= \delta_{ij} e^{pq}e_{pq}-4e_{ir}e_j \thinspace ^r.
\een
From (\ref{trace}) we deduce that
\ben
k_{00}= e_{rs} e^{rs}.
\een
and using (\ref{square}) we obtain 
\ben
{ 2\over 3}k^2 = 16 e_{ij} e^{jk} e_{ks} e^{si}- 4 (e_{ij} e^{ij})^2.
\een

Now using SO(5) transformations we can 
skew-diagonalize the two-form $e_{ij}$, that is choose
a basis in which the only non-vanishing components are $e_{12}=-e_{21}=e_1$
and $e_{34}=-e_{43}=e_2$.  In this basis, one finds that 
\ben
k^2= 24 (e_1^2 -e_2^2 )^2.
\een

The quantity $k_{mn}p^mq^n$ may be dealt with in a similar way. 
If $p^n$ is timelike we can 
choose a six-dimensional Lorentz frame in which
$p^m=(p^0,0,0,0,0,0)$. This choice allows us to use 
the $SO(5)$ freedom to skew diagonalize $e_{ij}$ as above.
A short calculation then gives
\ben
k_{mn}p^p q^n= 2p^0q^0(e_1^2 +e_2^2 -2{q^5 \over q^0}e_1e_2 ).
\een
Since $q^n$ is causal, $|{q^5 \over q^0} | \le 1$ and 
$e_1^2 + e^2_2 \ge 2 e_1e_2$,
we find the desired result, namely $k_{mn}p^m q^n \ge 0$.

Interestingly, $k^2$ vanishes
if and only if $e_{ij}$ determines a self-or anti self-dual
two-form in the four space orthogonal to its kernel.
We also see that $T_{mn}p^m p^n$ is strictly positive for timelike $p^n$ 
and hence taking the choice $p^n=(p^0,0,0,0,0,0)$ we conclude that 
$T_{00}$ is strictly positive.

\section{Exact Plane Wave solutions}

In this section we shall establish that plane
wave solutions of the linearized theory
are in fact {\sl exact} solutions of the full non-linear
equations of motion, a property which also holds in Born-Infeld theory
(compare \cite{GibbonsHerdeiro}) and classical general relativity.
We shall suppose that the metric $g_{mn}$ is flat, $g_{mn}=\eta_{mn}$,
although one might consider more general cases. 
We make the ansatz
\ben
H_{lmn}={H^0}_{lmn} f(l_n x^ n),
\een
where $H^0_{lmn}$ is a constant three-form, $f(u)$ is an arbitrary
function of its argument and the constant vector $l^n$ is null
\ben
\eta_{mn}l^ml^n=0 \label{nullness}.
\een
The closure condition becomes
\ben
l_{[l} H^0_{mno]} =0,
\een

Thus
\ben
\epsilon^{pqrstu}l_rH^0_{stu}=0 \label{C}.
\een 
If we assume that $l_m=(1,1,0,0,0,0)$ and let  greek indices
run from 2 to 5, we find that
\ben
H^0_{\alpha\beta\gamma}=0, \qquad H^0_{0\gamma \delta}=H^0_{1\gamma \delta}.
\een
This may be written covariantly
as
\ben
H^0_{mnp}= l_{[m} A^0_{np]}, \label{polarization}
\een
where $A^0_{np}$ is a constant polarization two-form.
It is determined only up to
\ben
A^0_{np} \rightarrow A^0_{np} + l_{[n} A^0_{p]},
\een
where $A^0_p$ is a constant one-form.

We now make the further assumption that

\ben l^m H_{mnp}=0.\label{additional}
\een

The reason we have to assume (\ref{additional}) is that  
quantities $H^0_{01\alpha}$ are not constrained
by  equation (\ref{C}). In order  to eliminate this freedom we are in effect
assumg that 
\ben
H^0_{01\alpha}=0.
\een

Having made this ansatz, i.e. assuming (\ref{polarization} ,
\ref{additional})
 it follows that
\ben
H_{lmn} H^{lmn}=0.
\een
Thus $Q=1$ and hence $C^{mn}=G^{mn}$.

It remains to solve the self-duality
condition. Now using (\ref{easymetric1}) we see that
five brane metric is of so-called Kerr-Schild form:
\ben
G^{mn}=\eta^{mn} -{\rm constant} f^2 l^m l^n.
\een
The constant is positive. To evaluate it we could introduce
a ( non-unique)  null vector $n^m$ , normalized so that
\ben
\eta_{mn}n^m l^n=-1.
\een

One then has
\ben
{\rm constant}= ({H^0}^2)_{mn} n^m n^n. 
\een

Because $l^m$ is null ( i.e. from (\ref{nullness}) )
it follows that 
\ben
G_m \thinspace ^q H_{npq}=H_{npm}.
\een

Thus the self-duality condition (\ref{easydual})
becomes
\ben
\star H_{lmn}=H_{lmn},
\een
which is the same as the condition for the linear theory.

Using Hooke's Law (\ref{Hooke}) we deduce that the energy momentum tensor
is given by
\ben
T^{mn}= {\rm constant } f^2   l^m l^n. \label{flu}
\een
Equation (\ref{flu}) is the energy momentum tensor of a 
fluid of energy density $\rm constant$
moving at the speed of light in the direction $l^m$.
This is often referred to as a null fluid.
We shall encounter null fluids again later when we    
examine the energy momentum tensor in the strong coupling limit.

\section{The $SO(5)$ covariant formalism and the  limiting field strength}

Born Infeld theory has a built in upper-bound
for the electric field strength. 
The string theoretic
interpretation is that when electric fields approach the
limiting field strength, copious pair production of
open string states occurs\cite{Burgess}\cite{BachasPoratti}. 
One may ask whether a similar
phenomenon takes place in M-Theory \cite{dutch}.

In Born-Infeld theory one must take care to express
the upper bound in terms of the correct variables.
The Lagrangian density is
\ben
L=1-\sqrt{1-{\bf E}^2+{\bf B}^2 - ({\bf E} \cdot {\bf B})^2 },
\een 
which shows that the electric field ${\bf E}$ cannot be too
big for fixed ${\bf B}$ since 
then the Lagrangian density becomes complex. 
However the Hamiltonian density \cite{Birula1} is
\ben
{\cal H}= 
\sqrt{ 1+{\bf B}^2 + {\bf D}^2 + ({\bf B} \times {\bf D})^2 } -1.
\een
This shows that there is no limit on either the magnetic induction
${\bf B}$ or electric inductions ${\bf D}$. 
However the dual Lagrangian \cite{Birula1} is
\ben
{\hat L}=\sqrt{1-{\bf H}^2+{\bf D}^2 -({\bf D} \cdot {\bf H})^2 }-1,
\een
which indicates that there is an upper bound on 
the magnetic intensity 
${\bf H}$,as there has to be, by electric-magnetic duality invariance.
This point is re-enforced by considering the dual 
Hamiltonian \cite{Birula1}
\ben
{\hat H}= 1-\sqrt{1- {\bf H}^2 -{\bf E}^2 
+ ({\bf H} \times {\bf E})^2 }.
\een
 
In the case of the M5-brane we have a similar 
range of possible choices of 
field variables. 
The fivebrane has a self-duality condition which means that 
if we define 
\ben 
B_{ij}=-{1\over 6}\epsilon _{ijklm}H^{klm},\ \  E_{ij}=H_{0ij}
\een
where $i,j,k,\ldots =1,2,\ldots ,5$,  then we can  
 express $E_{ij}$
in terms of $B_{ij}$ and  {\it vice versa}. Thus the 
energy density $T_{00}={\cal H}$ 
can be expressed in terms of either variable. To achieve this we will 
have in effect to solve the self-duality constraint. 
The self-duality condition on $H_{mnp}$ 
can be expressed as 
\ben
H^+_{mnp}=Q^{-1}h_{mnp}, \ H^-_{mnp}=2 Q^{-1}k_m{}^rh_{rnp}
\een
where 
$H^{\pm}_{mnp}={1\over2}(H_{mnp}\pm 
{1\over 3!}\epsilon_{mnprst}H^{rst})$. 
In the special frame used above one then finds that 
\ben
B_1={ e_1 \over 1-4(e_1^2 -e_2^2 )}, \space
B_2= {e_2 \over 1+4 (e_1^2 -e_2^2 ) }, 
\een
\ben
E_1={e_1 \over 1+ 4(e_1^2 -e_2^2 )}, 
\space
E_2={e_2 \over 1-4(e_1^2 -e_2^2 )}. \label{electric}
\een
Inverting these equations leads to
\ben
8e_1= { B_1 \over B^2_2 -B^2_1} \sqrt{ 1+16B_2^2} \Bigl [ \sqrt{ 1+ 16 B_2^2 }
- \sqrt{1+16 B_1^2 } \Bigr ],
\een
and 
\ben
8e_2= { B_2 \over B^2_1 -B^2_2} \sqrt{ 1+16B_1^2} 
\Bigl [ \sqrt{ 1+ 16 B_1^2 }
- \sqrt{1+16 B_2^2 } \Bigr ].
\een
Note that
\ben
E_1E_2= B_1 B_2.
\een

The required solutions of the self-duality constraint
are therefore
\ben
E_1= B_1 \sqrt {1+16B_2^2 \over 1+ 16 B_1^2 },
\space
E_2= B_2 \sqrt { 1+ 16 B_1^2 \over 1+ 16 B_2^2 },
\een
or inversely,
\ben
B_1 = E_1 \sqrt { 1- 16 E_2^2 \over 1- 16 E_1^2 }, 
\space
B_2= E_2 \sqrt{ 1-16 E_1^2 \over 1- 16 E_2^2 }\label{magnetic}.
\een

Expressing the energy momentum tensor in terms of
$h_{mnp}$ and using the relations of section four 
we find that 
\ben
T_{00}= 
{{1+ 16(e_1^2-e_2^2) + 8 ( e_1^2 + e_2^2 )} 
\over {1-16(e_2^2-e_1^2)^2} }
-1.
\een
Using the above equations we conclude that 
\ben
T_{00}= \sqrt{(1+16B_1^2)(1+ 16B_2^2)}-1 . \label{energydensity}
\een
The expression for the energy density (\ref{energydensity})
may be cast in the  SO(5) covariant form  
\ben 
{\cal H}= \sqrt {\det( \delta_{ij} + 4B_{ij})}-1 
= (\delta_{ij}+16 B_{ik} B_{jk} ) ^{{ 1\over 4}}-1.
\label{Hamilton}\een

A Hamiltonian density was derived from the action formulation
of the five-brane in \cite{SorokinHamiltonian}.
Our $\cal H$ coincides with that Hamiltonian density.

From (\ref{electric}) and (\ref{energydensity})
we have
\ben
E_1={ 1\over 16} {\partial {\cal H} \over \partial B_1}, \space
E_2={ 1\over 16} {\partial {\cal H} \over \partial B_2},
\een
which may be cast in the manifestly $SO(5)$-covariant form:
\ben
E_{ij} ={1 \over 16}{\partial {\cal H} \over \partial B_{ij
}} = { (1-8 {\rm Tr } B^2 ) B_{ij} + 16 B^3_{ij} \over \sqrt{ 1-8 {\rm
Tr } B^2 + (16)^2 W^2_i }} \label{elec},
\een
where
\ben
W_i={ 1 \over 8} \epsilon_{ijrst}B^{jr} B^{st}.
\een

Now the Legendre transform of the Hamiltonian 
density is
\ben
{\hat {\cal H}}= 1-\sqrt {(1-16 E_1^2) (1-16 E^2_1)}
= 1-\sqrt {\det ( \delta _{ij} + 4\sqrt{-1}E_{ij})},
\label{dualHamiltonian}\een
so that
\ben
{\hat {\cal H}} + {\cal H}= {1 \over 2} E_{ij} B^{ij}
\een
and
\ben
B_{ij}= { 1\over 16}{\partial 
{\hat {\cal H} } \over \partial E_{ij} }={ (1+8 {\rm Tr } E^2 ) E_{ij} + 16 E^3_{ij} \over \sqrt{ 1+8 {\rm
Tr } E^2 + (16)^2 U^2_i }} \label{mag} ,
\een
where
\ben
U_i={ 1 \over 8} \epsilon_{ijrst}E^{jr} E^{st}.
\een 
In our special frame this equation reduces to (\ref{magnetic}).
For a helpful review of the various formulations of
the 5-brane equations of motion which covers some of this material the reader
is referred  to \cite{Bandosetal}.

The  Hamiltonian density $\cal H$ is a well defined convex
function for all finite values of the magnetic induction
 $B_{ij}$. The Legendre transform maps
all of  $B_{ij}$ space in a one-one fashion
onto the open region
of $E_{ij}$ space for which the dual Hamiltonian
density ${\hat {\cal H}}$ is a well defined convex function.
That is for which  the matrix $\delta_{jk}-16E_{ji} E_{ki}$, 
which occurs in the dual Hamiltonian, 
is positive definite.
In our special frame, the allowed region is just $4|E_1| <1$ and
$4 E_2 <1$.

The Bianchi identities read 

\ben
\partial_iB_{ij}=0,\thinspace { \partial B_{ij} \over \partial t}
+{1 \over 2} \epsilon _{ijkrs} \partial_k E_{rs}=0 \label{bianchi}.
\een

Taking the  solution (\ref{elec}) for the self duality condition 
implies the equations of motion.

\section{Strong Coupling Limit}

 We are now going to discuss the strong coupling
limit of the theory. This is equivalent to 
taking the tension of the fivebrane to zero and may 
be thought of as a high energy limit. 
We shall find that the situation is 
similar to that of the strong coupling limit of 
Born-Infeld theory
in four spacetime dimensions which has been thoroughly 
investigated by
 Bia{\l}nicki-Birula and called by him UBI theory
\cite{Birula1, Birula2}.
In that case, the Lagrangian vanishes, but there is a well 
defined Hamiltonian.
Moreover the theory is Lorentz invariant and has the 
energy momentum tensor
of a null fluid. As a consequence 
there are infinitely many conserved currents in 
flat spacetime.
Because the invariants $F_{\mu \nu} F^{\mu \nu}$
and $ F_{\mu \nu} \star F_{\mu \nu } $                         
both vanish on shell one might speculate that the theory 
is quantum mechanically finite. As we shall see, there is 
just as much evidence to warrant a similar 
speculation about the M-Theory version. 

The limit we are considering differs from 
that discussed in \cite{dutch,Seiberg}. 

To proceed we introduce into the Hamiltonian a parameter $T$
with the dimensions of mass cubed. 
We then take the limit $T\downarrow 0$. The Hamiltonian density is
\ben
{\cal H}= T^2 \sqrt{ \det( 1+ {4 B_{ij} \over T }) }
 -T^2.
\een
Letting $T\downarrow 0$, we get the well defined limit
\ben
{\cal H} =16 |B_1B_2|.
\een

We now evaluate the energy momentum tensor 
in the strong coupling limit in terms of the variable $B_{ij}$. 
In this process we will find quantities that are non-vanishing 
only when one takes next to leading contribution in $T$. 
On finds that in the limit of small $T$,
\ben
e_1 \rightarrow { 2 \over T} {B_1B_2  \over B_1 + B_2 }
- { T \over 16}{ 1\over B_1} {B_1-B_2 \over B_1 + B_2 },
\een

\ben
e_2 \rightarrow { 2 \over T} {B_1B_2  \over B_1 + B_2 }
+ { T \over 16}{ 1\over B_1} {B_1-B_2 \over B_1 + B_2 }.
\een

It follows that
\ben
e_1^2 -e_2 ^2 \rightarrow { 1\over 4} {B_1-B_2 \over B_1 + B_2}.
\een

Reinstating the coupling constant $T$, the energy momentum tensor 
is given by

\ben
T_{mn}=-T^2 { m^2_{mn} \over Q} + T^2 g_{mn}.
\een

In the small $T$ limit the only non-vanishing components are 
\ben
T_{00} = 16|B_1B_2|,
\een

\ben
T_{55}= 16 |B_1B_2|,
\een

\ben
T_{05}=-16 B_1B_2.
\een

Thus 
\ben
T_{mm}
= {\cal H}l_ml_n,\label{covariant}
\een
with
\ben
l^m=(1,0,0,0,0,1),
\een
and  
\ben
l^ml^n g_{mn}=0\label{null}, 
\een
We have been working in a local frame. However, 
equations (\ref{covariant}) and (\ref{null}) are manifestly covariant 
and hence hold in a general frame.
In particular we have in general, that  the null vector $l^m$
is given by
\ben
l^m=(1,n^i),
\een
where $n^i$ is a unit vector in the direction of the Poynting vector
$T^{0i}$ and therefore 
\ben
T^{om}= {\cal H} l^m.
\een

From (\ref{covariant}) we see that 
the trace of the energy momentum tensor vanishes
\ben
T^m_m=0.
\een
Thus we have shown that the theory is Weyl invariant in the strong 
coupling limit. The energy momentum tensor (\ref{covariant}) 
is 
the same form as the energy momentum tensor of the 
strong coupling limit
of Born-Infeld theory, UBI   theory.

Because the energy momentum tensor has the form of a 
null fluid ( see \cite{Birula1, Birula2}) one may easily check that 
one has infinitely many conservation laws. 
The conservation of energy equation $\partial_t T^{00} +\partial _iT^{0i}=0$
may be written covariantly as
\ben
\nabla _m({\cal H} l^m)=0. \label{entropy}
\een                       
The remaining conservation law implies that
\ben
l^n\nabla _n l^l=0. \label{geodesic}.
\een
From (\ref{geodesic})it follows  that the integral curves tangent to
$l^m$, that is the solutions of
\ben
{dx^m \over d \lambda}=l^m,
\een
are null geodesics with affine parameter $\lambda$.
In fluid dynamic language (\ref{entropy}) corresponds
to the conservation of entropy, sometimes thought
of as `photon" number. The integral curves are thought of as the
world lines of the fluid. In the case of UBI theory
\cite{Birula1, Birula2} it turns out \cite{Gibbonsunpub}
cf. \cite{Yi} that the fluid may be thought of as
a gas of massless  Schild strings \cite{Schild}.
In fluid dynamic language the world-sheets of the strings
are the histories of magnetic field lines which are swept
out with velocity $l^m$. We shall turn later
to the possible fluid interpretation in the M5-brane case.

It follows from (\ref{entropy}) and (\ref{geodesic})
that the tensors
\ben
T^{m_1m_2\dots m_k}= {\cal
H}
 l^{m_1} l^{m_2}\dots l^{m_k}
\een
satisfy
\ben
\nabla _m T^{m m_2 \dots m_k}=0, \label{divergence2}
\een
for any positive integer $k$.

In flat spacetime the divergence
identities (\ref{divergence2} )
 may be integrated over space to give rise to infinitely many conservation laws.

In order to cast the limiting  equations
of motion in a manifestly $SO(5)$ covariant form, we recall from
(\ref{elec}) and (\ref{bianchi})
that
\ben
{ \partial B^{ij} \over \partial t }+{ 1\over 2} 
\epsilon ^{ijrst} \partial_r
 \Bigl \{ 
{ (B-8({\rm Tr} B^2) B +16  B^3 )_{st}
\over \sqrt{ 1- 8 {\rm Tr }B^2 +16^2 W^2_i} } 
 \Bigr \}  =0.\label{eqnmot}
\een

One may now re-instate $T$ and then take the limit $T\downarrow 0$ 
in (\ref{eqnmot}) to obtain

\ben
{ \partial B^{ij} \over \partial t }+{ 1\over 2} 
\epsilon ^{ijrst} \partial_r
 \Bigl \{ 
{ (-{ 1 \over 2} ({\rm Tr} B^2) B + B^3 )_{st}
\over \sqrt{W^2_i} } 
 \Bigr \}  =0.
\een
 
Using the identity
\ben
(-{\rm Tr} B^2 B+2 B^3 )^{ij}=\epsilon^{ijklm}W_kB_{lm},
\een
we may show that  in the limit, electric field becomes
\ben
E^{ij}= { 1\over 2} \epsilon ^ {ijrst} { W_r \over \sqrt{W^2_k} }
B_{st} ,\een
and the field equation becomes
\ben
{\partial B_{ij} \over \partial t}+ { 1\over 4}
\epsilon_{ij}\thinspace^{klm} \partial _k \epsilon_{lmrst}{ W^r B^{st}
\over \sqrt{W^2_p}}. 
\een

This is very similar in form  to the Born-Infeld case \cite{Birula1, Birula2}.
One  may also give an $SO(5,1)$ covariant formulation of the 5-brane
in the $T\downarrow 0$ limit. 
It is sufficient to derive the
self-duality
conditions in this limit as it, together with the Bianchi Identity,
imply the field equations.

By multiplying through by $m^{-1}$, the self-duality condition $*
H_{abc}= Q^{-1}(m^2)_a{}^e H_{ebc}$ can be reexpressed as 
\ben
(m^{-1})_a{}^e * H_{ebc}= Q^{-1}(m)_a{}^e H_{ebc}
\een
or 
\ben(1+2k)_a{}^e * H_{ebc}= Q^{-1}(1-2k)_a{}^e H_{ebc}
\een
Taking the limit and using the equation  for $k$ in terms of $H^2$ we
find  that the self-duality condition can be expressed covariantly
as 
\ben(H^2)_a{}^e* H_{ebc}=({1\over 3} H^2\delta _a^e - (H^2)_a{}^e) 
H_{ebc}
\een

In view of the interpretation of the strong coupling limits
of Born-Infeld theory as string fluids \cite{Yi}
, it is  of interest
to consider the detailed structure of the three-form
H. Point-wise, i.e. locally, it may be expressed in the  general
case  as
\ben
H=E_1 dt \wedge dx^1 \wedge dx^2 + E_2 dt \wedge dx^3 \wedge dx ^4 -B_2 dx^5 \wedge dx^1 \wedge dx^2 -B_1 dx^5 \wedge dx^3 \wedge dx^4.
\een
\ben
\star H=E_1 dx^3  \wedge dx^4 \wedge dx^5 + E_2 x^1 \wedge dx^2 \wedge dx ^5 +B_2 dt \wedge dx^3 \wedge dx^4 +B_1 dt\wedge dx^1 \wedge dx^2.
\een
In the strong coupling limit we get the well-defined limits
\ben
H=(dt-dx^5) \wedge (B_2 dx^1 \wedge dx^2 + B_1 dx^4 \wedge dx^5 ),
\een
\ben
* H=(dt-dx^5) \wedge (B_2 dx^2 \wedge dx^3 + B_1 dx^1   \wedge dx^2 ).
\een
By contrast since
\ben
h=e_1(dt \wedge dx^1 \wedge dx^2 -dx^5 \wedge dx^3 \wedge dx^4 ) + e_2 (dt \wedge dx^3 \wedge dx^4 - dx^5 \wedge dx^1 \wedge dx^2 ),
\een
the three-form $h_{lmn}$ diverges in the limit of small $T$
\ben
h \rightarrow { 2 \over T} {B_1 B_2 \over B_1+ B_2} (dt-dx^5) (dx^1 \wedge dx^2 + dx^3 \wedge dx^5 ).
\een

We see that in the limit
$H$ and $\star H$ contain a one dimensional null factor 
$dt-dx^5=-l_mdx^m$, with

\ben
H_{lmn} l^n=0 ,\thinspace \thinspace  
\star H_{lmn} l^n=0.
\een

It is tempting to give the stress tensor
an interpretation in terms of a fluid of 
$p$-branes. The null vector $l^m$ clearly
provides us with the velocity of the
putative  fluid
and since it is null, we should 
expect a null fluid analogous to 
the gas of null or Schild strings \cite{Schild}  considered
 in \cite{Letelier} and elaborated in 
\cite{Stachel1} \cite{Stachel2} in four dimensions. 
In the case of UBI theory \cite{Birula1,Birula2}
one has the condition
\ben
\sqrt{\det ( F_{\mu \nu } ) }=
{1\over 4} F_{\mu \nu} \star F^{\mu \nu}= {\bf E }\cdot {\bf B}=0 \label{simple1}. 
\een
This is equivalent to the simplicity condition
\ben
F \wedge F=0, \label{simple2}
\een
which implies the existence of one-forms $a_\mu $ and $b_\mu$ such that
\ben
F_{\mu \nu} = a_\mu b_\nu -a_\nu b_\mu.
\een                                   
The co-vectors $a_\mu$ and $b_\mu$ are not unique;
one may choose  them such that $a_\mu b^\mu$=0.
Thus, in  the tangent space at  each point of spacetime,
the two-form $F_{\mu \nu}$ is tangent to the two-plane
spanned by $a_\mu$ and $b_\mu$. The equations of motion
imply the integrability condition that these two-planes mesh
together to provide a foliation of spacetime by two-surfaces.
If
\ben
F_{\mu \nu} F^{\mu \nu}=0 \label{zero},
\een
then the  two-plane is null, and one of the co-vectors,
call it $a_\mu$, is null and the other spacelike.
Therefore
\ben 
F_{\mu \nu} = l_\mu b_\nu - l_\nu b_\mu.
\een
The null vector $l_\mu$ satisfies
\ben
F_{\mu \nu} l^\nu=0.
\een
If (\ref{zero}) then the principal  null directions
of the two-form $F_{\mu \nu}$ 
coincide and may be identified with $l^\mu$.    
One may regard  $b_\mu$ as the 
(un-normalised) tangent vector to the  magnetic field
lines and  obtain in this way a picture of the
solutions of classical Born-Infeld field theory, 
subject to the simplicity
constraint (\ref{simple1}) or (\ref{simple2}),
as a fluid of strings. This viewpoint is
similar to that
originally  envisaged by Nielsen and Olesen \cite{NielsenOlesen}.

In the case of the M5-brane, we have a three-form and one might have 
thought that some sort of null membrane is involved.
However although the three-forms $H$ and and $\star H$ both 
have a null factor neither quotient two form. that is
\ben
{\rm neither }\thinspace \thinspace B_2 dx^1 \wedge dx^2 + B_1 dx^4 \wedge dx^5 
\thinspace \thinspace  {\rm nor  }\thinspace  \thinspace B_2 dx^2 \wedge dx^3 + B_1 dx^1   \wedge dx^2
\een
is simple. Therefore no obvious membrane world volume is picked out.

We observe that this limiting theory admits an 
infinite number of conserved 
charges that carry Lorentz indices. Presumably (c.f. \cite{ColemanMandula})
the scattering of particle-like excitations is 
trivial in this limit.
It is not obvious whether this is true for 2-branes and 5-branes
since they have internal degrees of freedom associated to their world
volumes.

\section{Acknowledgement}

We should like to thank the hospitality
of the Theory Group at CERN and IHP, where
part of this work was carried out. 
This work was supported in part by the two EU networks entitled  "On 
Integrability, Nonperturbative effects, and Symmetry in Quantum Field 
Theory"  (FMRX-CT96-0012) and "Superstrings" (HPRN-CT-2000-00122). It was also 
supported by the PPARC special grant PPA/G/S/1998/0061.

\end{document}